\documentstyle[aps,prl,preprint]{revtex}

\begin{document}
\draft
\title{Evidence for charge-flux duality near the quantum Hall liquid to
insulator transition}
\author{D. Shahar, D.C. Tsui and M. Shayegan}
\address{Department of Electrical Engineering, Princeton University,
 Princeton New Jersey, 08544}
\author{E. Shimshoni\cite{byline1} and S.L. Sondhi\cite{byline2}}
\address{Department of Physics, University of Illinois at Urbana-Champaign,
Urbana IL 61801}

\date{\today}

\maketitle
\begin{abstract}
We examine the longitudinal, non-linear, current-voltage characteristics near
the quantum Hall liquid to insulator transition and show that a simple mapping
exists between the characteristics on the quantum Hall side and those on
the insulating side of the transition. More precisely, at filling factors
related by the law of corresponding states the current and voltage simply
trade places. We interpret these observations as evidence for the existence,
in the composite boson description, of charge-flux duality near disorder
dominated transitions in quantum Hall systems.
\end{abstract}

%\twocolumn

The quantum Hall effect (QHE) has long been understood to involve
the intertwining of the elegant physics of the ideal states with
the equally elegant physics of localization.
The precise nature of the intertwining has
been somewhat controversial, for it involves solving a problem with
interactions, fractional statistics and disorder---a somewhat
formidable mix. Nevertheless, considerable theoretical and experimental
progress had been made in recent years in understanding the phase diagram
of QH states and the transitions between them.

On the theoretical front, Kivelson, Lee and Zhang (KLZ) \cite{KLZ} used a
flux attachment (Chern-Simons) transformation \cite{ZHK}, to map the
two dimensional electron system (2DES) at a magnetic field
($B$) onto a bosonic system under a
different field $B_{eff}$. The advantage of this mapping is clear if one
considers the ``magic'' Landau level filling factors ($\nu$'s)
where the fractional quantum Hall effect (FQHE) liquid states are observed.
At these $\nu$'s the Chern Simons gauge field cancels, on average,
the externally applied $B$, and the composite bosons (CB) experience
a vanishing $B_{eff}$. The incompressible FQHE states then arise
as a result of the formation of a
Bose-condensed, superconducting, state of the CB's.

At $\nu$'s other than the magic $\nu$'s, the
cancellation of the external $B$ is not exact and, according to this
bosonic picture, vortices are created in the CB condensate. For small
deviations from magic $\nu$'s the vortex density is small, they
are localized by disorder and do not contribute to the
long wavelength electrical response.
When the deviation from the magic $\nu$'s becomes sufficiently large
the superconductivity of the bosons is destroyed by the excess magnetic
field and they turn insulating. This results in a transition to an
insulating, or a different quantum Hall, state.
Utilizing this bosonic formulation, KLZ derived a ``law of
corresponding states" which relates the different QH states and allows
construction of a phase diagram for the QH system in the
disorder--$B$ field plane. This law implies that all continuous
phase transitions between QH states and between QH states and
insulators are in the same universality class when described in terms of
the composite bosons. Hence, they should be characterized by the same
exponents and by universal critical resistivities that are related by the
corresponding states formulae \cite{fn1}.

Remarkably, much of KLZ's account of the phase diagram and transitions
is in agreement with experiments. Most of the observed transitions
follow the KLZ phase diagram \cite{fn2} and
the scaling behavior found near the transitions
between adjacent
QHE states\cite{wei,engel}, as well as the transitions from the QHE states
to an insulating phase \cite{Wong95,Shahar:PUthesis}
indeed appears to be independent of the transition involved. Finally,
Shahar et al \cite{DShahar95} found well--defined values for the critical
resistivities at transitions from the $\nu=1$ and $\nu=1/3$
QH liquids to the
proximate insulators, as predicted by
the correspondence rules.
Similar results have also been reported by Wong et al \cite{Wong95}.

In this letter we report the results of a set of measurements
that extend the framework developed thus far in an unexpected direction.
We have studied both the longitudinal ($V_{xx}$) and
Hall ($V_{xy}$) current--voltage ($I-V$)
characteristics near the QH to Insulator transition
and find that a) the $I-V_{xx}$ curves are strongly non--linear,
b) they exhibit a unique {\em reflection} symmetry between certain
traces in the
QH state and in the insulating phase, and c) at the same $B$
and $I$ values where the $I-V_{xx}$ traces are stronly non--linear,
the $I-V_{xy}$ curves are {\em linear} and virtually $B$ independent.
We will argue below that, taken together, these observations suggest
that the CB action exhibits a form of charge-flux or
electric--magnetic duality \cite{duality}
{\em and} that Hall resistance of the CBs vanishes on both sides of the
transition.

\noindent{\bf Longitudinal Response and Reflection Symmetry:}
The results presented in this letter were obtained from a
high mobility ($\mu=5.5 \cdot 10^{5}$ cm$^2$/Vsec), low density
($n=6.5\cdot 10^{10}$ cm$^{-2}$) MBE grown GaAs/Al$_{x}$Ga$_{1-x}$As
heterostructure.
A typical $B$ trace of the diagonal resistivity, $\rho_{xx}$, for this
sample is plotted in Fig. 1. Attesting to the high quality of this sample
is the set of fractional quantum Hall effect (FQHE) states that
includes
the $\nu=3/4$, $2/3$, $3/5$, $2/5$ and $1/3$ states which are manifested
as minima in the $\rho_{xx}$ trace. The QHE series terminates at high $B$
with a transition to an insulating phase. The transition point, $B_{c}$,
is identified\cite{DShahar95} with the common crossing point of the $B$
field traces of $\rho_{xx}$ taken at several $T$'s between $26$ and $88$ mK
(c.f. the inset of Fig. 1). A systematic study of this transition was
presented in Ref.\cite{DShahar95}.

In Fig 2a we present a selected set of $I-V_{xx}$
traces obtained from our sample at $T=21$ mK, taken
at various $B$ values in the vicinity of $B_{c}$.
With the exception of the trace taken closest to $B_{c}$,
the $I-V_{xx}$ curves are markedly non-linear both in the
$\nu=1/3$ FQHE regime (short-dashed lines) and in the insulating phase
(solid lines).
Deep in the $\nu=1/3$ FQHE regime ($B=8.5$ T), strong superlinear
behavior is observed, typical of a dissipationless QHE state. As we approach
$B_{c}$, this superlinearity becomes less pronounced,
and the curve taken closest $B_{c}$ (long-dashed line)
appears linear over the
entire voltage range, within experimental error. At higher $B$, as we
enter the insulating phase, the $I-V_{xx}$ curves (solid
lines) change in nature and are now sublinear, reflecting the
insulating nature of the conduction process. The sublinearity increases
as we go deeper into the insulating phase.
In this figure, we normalized the I and V units such that
the dashed line, which is the $I-V$ curve closest to $B_{c}$,
has unity slope. This amounts to defining the resistivity units in
terms of the critical resistivity at the transition point, which was
shown to be close to $h/e^{2}$\cite{DShahar95}.
For this sample it is $23$ k$\Omega$.

A striking feature in the set of $I-V_{xx}$ curves presented in Fig. 2a is
the existence of an apparent reflection symmetry about the linear $I-V_{xx}$
curve taken at $B_{c}$. To illustrate this symmetry we plot, in Fig.
2b, the same $I-V$ curves as in Fig. 2a, but with the current and
voltage coordinates of the curves taken in the FQHE regime exchanged.
Remarkably, the overlap between some pairs of traces is almost perfect,
to within experimental error. This matching is
maintained rather far from the transition point, over a range of $1.6$ T.
Similar symmetry is observed also for the $\nu=1$ to insulator
transition, but over a rather narrower range in $B$ of $0.1$
T\cite{Shahar:LongIV}.
On a purely phenomenological level, the correspondence of the $I-V$ traces on
both sides of the transition indicates that at least some
of the special correlations that are responsible for the FQHE state survive
the transition to the insulating phase. In other words, there exists
a relation between the physical mechanism responsible for the
non-linear $I-V$ at a given $B$ in the FQHE liquid regime
and that at a corresponding $B$ value in the insulating phase. We will
argue below that a plausible explanation for this experimental finding
is the existence of a charge-flux duality near the transition.

In order to investigate the nature of this correspondence, it is
necessary to identify whether a systematic relation exists between
the $B$ values of the matched pairs of $I-V_{xx}$ traces in Fig. 2b.
In Fig. 3, we summarize the $\Delta B$ (Fig. 3a) and $\Delta \nu$
(Fig. 3b) values for each one of the $I-V_{xx}$ traces in Fig. 2.
The abscissa designates the pairs, which are identified by a number
in Fig 2b and 3, smaller numbers denote pairs closer to the transition; in
the latter, solid circles designate $I-V_{xx}$'s taken in the $1/3$ FQHE
state and empty
circles those taken in the insulator. Evidently, the matched pairs are
neither the same $\Delta B$ from the transition, nor the same $\Delta
\nu$.

A clue to the correspondence is the interpretation of the reflection
symmetry as the statement that the current and voltage exchange
roles across the transition. The phenomenology of superconductors in the
vortex state, where vortex conduction leads to electrical resistance,
suggests that this is a consequence of an interchange
between the roles of bosons and vortices, i.e. of a charge-flux duality.
This notion can be made more concrete in the framework of the composite boson
description. Recall that the state at a generic filling factor contains
both bosons and vortices---in the QH phase the bosons superconduct while
the vortices are localized while in the insulator the vortices superconduct
and the bosons are localized. Hence the natural candidate for the dual
of a filling with $N_b$ bosons and $N_v$ vortices is the filling factor
with $N_v$ bosons and $N_b$ vortices.

In the vicinity of the $\nu=1/3$ to insulator transition, the composite
bosons ``carry'' three quanta of flux and hence at filling $\nu$ the
state of a system with $N_b=N$ bosons has $N_v= N(\nu^{-1} - 3)$ vortices.
It follows from this that its dual filling is
$\nu_d = (1-3\nu)/(\nu + 3(1-3\nu))$.
A different way to express this, and one which makes contact with the
composite fermion description, is that the auxiliary fillings
$\nu'=\frac{\nu}{1-2\nu}$, are related as
$\nu'=\nu'_{c}+\Delta\nu'$ and $\nu'_d =\nu'_{c}-\Delta\nu'$. Or, that
duality for composite bosons is the same as particle-hole symmetry for
their fermionic cousins. Fig 3c shows that the correspondence in terms
of the auxiliary fillings describes our data significantly better than
the previous attempts.

\noindent
{\bf Boson Resistivities and Hall voltages:} Thus far we have introduced the
notion of a duality to summarize the experimental observation that
current and longitudinal voltage appear to trade places at fillings where
the numbers
of bosons and vortices trade places in the composite boson description.
Evidently, the exchange in the numbers of bosons and vortices can lead
to a symmetry of the transport data only if their {\em dynamics} are
also identical.
While theoretical arguments to this effect, i.e. that the theory in the
critical region exhibits self-duality \cite{fn3}, will be presented elsewhere
\cite{shimshoni}, here we proceed more phenomenologically by focusing
temporarily on the {\em linear} response
and invoking the notion of boson resistivities \cite{KLZ}.

For the transition that we
have studied, they are defined by the decomposition ($h/e^2=1$):
\begin{equation}
\left(
\begin{array}{rr}
\rho_{xx}&\rho_{xy}\\
-\rho_{xy}&\rho_{xx}
\end{array}
\right)
=
\left(
\begin{array}{rr}
\rho_{xx}^b&\rho_{xy}^b\\
-\rho_{xy}^b&\rho_{xx}^b
\end{array}
\right)
+
\left(
\begin{array}{rr}
0&3\\
-3&0
\end{array}
\right)
\end{equation}
i.e. the observed resistivities are the resistivities of
the composite bosons corrected for the fact that they carry flux, which
produces a purely Hall voltage. Duality for the bosons would imply that
their resistivity tensor at $\nu$ is the same as the resistivity tensor
of the vortices at $\nu_d$ which in turn is the same as the
conductivity tensor of the bosons at $\nu_d$. Now, the
observed reflection symmetry of the $I-V_{xx}$ curves implies that
$\rho_{xx}(\nu) = 1/\rho_{xx}(\nu_d)$ \cite{fn:OhmicDuality}.
This implies that the bosons exhibit the
restricted duality $\rho_{xx}^b(\nu) =  1/\rho_{xx}^b(\nu_d)$
{\em and} $\rho_{xy}^b = 0$ everywhere in the vicinity of the
transition, including the insulating side. While this constraint is
limited to the linear response of the bosons, it is also hard to see
how a current-voltage duality could arise in the
longitudinal response (alone) beyond the ohmic
region if the bosons exhibited a non-zero Hall response.
This then
suggests that the observed
Hall voltage is a ``Faraday effect'' in that it
comes only from the ``flux'' carried by the bosons. This
chain of inferences can be used to predict the behavior of
$\rho_{xy}$, the Hall resistivity of the electrons:
It implies that a) $\rho_{xy} = 3$ everywhere
in the transition region and b) the Hall voltage should be linear
with driving current when the longitudinal voltage is {\em not}.

To test these predictions we plot, in Fig. 4,
$V_{xy}$ vs $I$ traces taken at $T=21$ mK and at $B$
intervals  of $0.1$ T starting in the $\nu=1/3$ FQHE
phase ($B=8.7$ T) and ending in the insulating phase ($B=9.7$ T).
Remarkably, these measurements are in accordance with both predictions.
In stark contrast to the longitudinal $I-V$s taken in the same
$B$ range (Fig. 2a), these curves are virtually $B$ independent,
and are all rather linear with slope $3 h/e^2$, within experimental error.

\noindent
{\bf Summary:} Our purpose in this letter has been primarily to
report two striking experimental observations on the transition region:
the reflection symmetry of the non-linear longitudinal $I-V$'s, and the
linearity and constancy of the Hall voltage for the same range of
currents and $B$ fields.
The theoretical ideas that we have discussed
seem to be plausible inferences from the data and certainly seem to tie the
different measurements together in a very economical way; indeed, they led
us to some of the measurements.
We will discuss elsewhere \cite{shimshoni} the precise
connection of these ideas to earlier work on duality transformations
in quantum Hall systems and on a ``semi-circle law''
for the conductivity tensor \cite{ruzin} which is equivalent to our
statement that $\rho_{xy} = 3$. Here we would like to note that the
existence of a symmetry in the non-linear response was not anticipated,
and raises issues concerning the role of dissipation near
quantum critical points that do not arise in treatments of the linear
response alone \cite{nonlinear} which will need to need to be addressed
in future work. Finally, we believe that the duality discussed here for
the QH/insulator transitions is present also at transitions between QH states
(e.g. between $\nu=1$ and $\nu=2$) but detecting its presence will require
extracting the response of the ``component'' undergoing a QH to Insulator
transition (the upper Landau level) from the measured transport data
beyond the linear regime \cite{shimshoni}.

\noindent
{\bf Acknowledgments:} We thank L. W. Engel, M. P. A. Fisher, S. M. Girvin,
S. A. Kivelson and M. Stone for instructive discussions.
This work has been supported by the NSF and the Beckman Foundation.

%%%%%% Fig 1 %%%%%%
\begin{figure}
\caption{A typical $B$ trace of $\rho_{xx}$ for our sample.
 The inset focuses on the transition region near $B_{c}=9.1$ T. The
 temperatures for the traces are 26, 36, 48, 65 and 88 mK.}
\end{figure}

%%%%%% Fig 2 %%%%%%
\begin{figure}
\caption{(a) $I$ vs $V_{xx}$ at several $B$'s near $B_{c}$. Solid
lines are in the insulating phase and dotted lines are in the
$\nu=1/3$ FQHE liquid. The $B$ values are 8.5, 8.6, 8.7, 8.9, 9, 9.1
(dashed line), 9.2, 9.3, 9.6, 9.8, and 10.1 T. (b) The same data as
in (a) but with the $I$ and $V_{xx}$ coordinates of the traces in the
$\nu=1/3$ FQHE state (dotted lines) exchanged. The numbers are used to
identify the pairs.}
\end{figure}

%%%%%% Fig 3 %%%%%%
\begin{figure}
\caption{(a) $\Delta B=B-B_{c}$, (b) $\Delta \nu=\nu-\nu_{c}$ and
(c) $\Delta \nu'=\nu'-\nu'_{c}$  (see text) for each pair
of matched $I-V_{xx}$'s in Fig. 2b vs the pair number. Pairs with
smaller numbers are closer to the transition.
}
\end{figure}

%%%%%% Fig 4 %%%%%%
\begin{figure}
\caption{$V_{xy}$ vs $I$ traces
near the transition, taken at 0.1 T steps in the $B$ range
indicated.}
\end{figure}


\begin{references}
\bibitem{fn1} This work involves treating the disorder within the RPA which
can be problematic, especially near critical points. However, as we note,
it appears that the current experiments are in excellent agreement with
theory at this level of approximation---whether this agreement is more
fundamental is not known.
\bibitem{fn:OhmicDuality}  Initial investigation of
duality in the (strictly) linear response regime shows that indeed
$\rho_{xx}(\nu) = 1/\rho_{xx}(\nu_d)$ close to the transition
point, with systematic deviations from this equality as we go farther
from the transition. A detailed study of duality in the ohmic regime
will be the subject of a future publication.
\bibitem{fn3} By self-duality we mean that the quantum action in terms
of the particles is identical to the action in terms of their duals (the
vortices) at the dual values of the parameters (here, filling factor).
This usage is intended to distinguish our problem from cases where
there is a dual description but it doesn't resemble the original one.
Exactly at the critical point the system is said to be self-dual in a
stronger sense namely the dual parameters are also identical to the
original ones---i.e. duality is actually a symmetry.
\end{references}
\end{document}